\documentclass[prl,aps,twocolumn,superscriptaddress,showpacs,floatfix]{revtex4}
\usepackage{graphicx}
\usepackage{subfigure}
\usepackage{amsmath, amssymb}
\begin{document}
\DeclareGraphicsExtensions{.eps}
\title{The Non-equilibrium Behavior of Fluctuation Induced Forces}
\author{David S. Dean}
\address{Kavli Institute for Theoretical Physics, University of California Santa Barbara, CA 93106, U.S.A.}
\address{Laboratoire de Physique Th\'eorique, IRSAMC, Universite Paul Sabatier, 118 Route de Narbonne, 31062 Toulouse Cedex 4, France}
\author{Ajay Gopinathan}
\address{Kavli Institute for Theoretical Physics, University of California Santa Barbara, CA 93106, U.S.A.}
\address{School of Natural Sciences,
University of California,
Merced, CA 95344, U.S.A.}

\begin{abstract}
While techniques to compute thermal fluctuation induced, or pseudo-Casimir, forces in equilibrium systems are well established, the same is not true for non-equilibrium cases. We present a general formalism that allows us to unambiguously compute non-equilibrium fluctuation induced forces  by specifying the energy of interaction of the fluctuating fields with the boundaries. For a general class of classical fields with  dissipative dynamics, we derive a very general relation between the Laplace transform of the time-dependent  force and the static partition function for  a related problem with a different Hamiltonian. In particular, we demonstrate the power of our approach by computing, for the first time, the explicit time dependence of the non-equilibrium pseudo-Casimir force induced between two parallel plates, upon a sudden change in the temperature of the system. We also show how our results can be used to determine the steady-state behavior of the non-equilibrium force in systems where the fluctuations are driven by colored noise.
\end{abstract}

\maketitle

The Casimir effect arises when the fluctuations of a quantum or classical field are constrained by
the presence of surfaces or objects placed in the field  \cite{kar1999,mos1997,mil2001}. While the standard Casimir effect refers to interactions arising from constrained fluctuations of an electromagnetic field, a similar effect that arises due to constrained {\em thermal} fluctuations in systems with long range correlations, such as critical fluids, smectic manifolds or liquid crystals, is termed the pseudo-Casimir effect. 
The simplest system where the so called pseudo-Casimir effect arises is  the classical  free scalar field
theory, and long range pseudo-Casimir forces arise when the field theory is massless, {\em i.e.} where the Hamiltonian of the system is given by:
\begin{equation}
H = {1\over 2} \int d{\bf x} \ \left[\nabla\phi({\bf x})\right]^2.\label{ham}
\end{equation}
 The  fluctuating field $\phi$ can describe the order parameter for critical systems, such as binary liquids at the critical point \cite{fi1978}, and the occurrence of a pseudo-Casimir interaction in such a system has recently been confirmed experimentally \cite{he2008}. The field $\phi$ can also represent the phase of the complex order parameter in a superfluid state, where pseudo-Casimir forces can lead to the thinning of superfluid $^4He$ films \cite{zan2004}.   Free vectorial field theories describing liquid crystal systems also exhibit pseudo-Casimir type interactions between the surfaces confining the system \cite{ad1991}. The interaction induced between surfaces or objects immersed in such systems can be considered to be due to the 
imposition of boundary conditions on the field or due to an energy of interaction of the surfaces or objects with the field. If, for instance, we consider the surfaces to be two parallel plates and  specify the boundary conditions for the field on the two plates ({\it e.g.} Dirichlet, Neumann or Robin), the pseudo-Casimir interaction between the plates can be computed from the free energy using standard techniques  {\em if} the system is at thermal equilibrium \cite{kar1999}. The same techniques {\em cannot},  however, be used to compute the pseudo-Casimir force if the system is driven out of equilibrium, for example by  a sudden change in temperature or by colored noise forcing. Since many experimental and naturally occurring systems, where such fluctuation induced forces are important ({\it e.g.} between inclusions in cell membranes), are actually out of equilibrium, it is essential to have a well-defined, unambiguous way to compute non-equilibrium pseudo-Casimir forces. One of the principal problems when analyzing the 
non-equilibrium pseudo-Casimir effect, say, for parallel plates, is to obtain a correct expression for the force between the two plates.
In previous studies, the stress tensor has been used to study both
the dynamical behavior of the force \cite{ba2003,na2004,ga2006} and the force fluctuations in equilibrium \cite{ba2002}. However, only the average force in equilibrium can be strictly computed using the stress tensor. Results using the stress tensor may, however, be reliable for situations close to equilibrium \cite{ga2006}. Another approach is to construct a 
model with a specified non equilibrium dynamics and to specify by hand the force at the wall. For example,
in \cite{br2007}, the dynamical field was related to a particle density and the local pressure
on the wall is then given by the ideal gas form via kinetic reasoning.
In this letter, rather than imposing boundary conditions on the field we specify the energy of interaction of the wall with the field. In this way we can write down the instantaneous force on the wall unambiguously. Our novel approach can be used to recover results for the usual  boundary conditions employed in studies of the 
pseudo-Casimir interaction by taking the appropriate limit but is, in fact, extremely general. We believe that such a microscopic  approach is required to obtain meaningful results  for the pseudo-Casimir force out of equilibrium.

We commence by considering the most general case of a free field theory where the Hamiltonian can be
written in terms of a general quadratic Hamiltonian. 
\begin{equation}
H ={1\over 2} \int d{\bf x} d{\bf x}' \phi({\bf x})\Delta({\bf x}, {\bf x}', l)\phi({\bf x}')\label{eqH},
\end{equation}
where $\Delta$ is a self-adoint operator {\em i.e.}~$\Delta({\bf x},{\bf x}')=\Delta({\bf x}',{\bf x})$.
Here $l$ represents any suitable free parameter in the problem but, for concreteness, it could be
the separation of two parallel plates which interact with the field. Now, if we choose
\begin{equation}
\Delta({\bf x}, {\bf x}',l) = \left[- \nabla^2 + \delta(z) c + \delta(z-l) c\right]\delta({\bf x}-{\bf x'}),
\end{equation}
then this corresponds to a free field theory where the fluctuations of the field $\phi$ are suppressed at both plates (at $z=0$ and at $z=l$). Clearly, when $c\to \infty$ one
will obtain Dirichlet boundary conditions at the two plates. The instantaneous generalized force acting on the plate at $z=l$ is then given by
\begin{equation}
F_l = -{\partial H\over \partial l} =-{1\over 2} 
\int d{\bf x} d{\bf x}' \phi({\bf x}){\partial \over \partial l} \Delta({\bf x}, {\bf x}', l)\phi({\bf x}').\label{eqforce}
\end{equation}
The equilibrium value of this force can also be written in the familiar form
\begin{equation}
\langle F_l\rangle = T{\partial\over \partial l}\ln(Z(\Delta)) \label{eqfe}
\end{equation}
where 
$
Z(\Delta) = \int d[\phi] \exp(-\beta H[\Delta] )
$
with $H$ as defined in Eq.(\ref{eqH}) and where $\beta = 1/T$ is the inverse temperature. 
Alternatively the force can be expressed via Eq.(\ref{eqforce}) as
\begin{equation}
\langle F_l \rangle =-{T\over 2} 
\int d{\bf x} d{\bf x}' \ \left[{\partial \over \partial l} \Delta({\bf x}, {\bf x}', l)\right]
\Delta^{-1}({\bf x}, {\bf x}', l)
\end{equation}

We now consider the dynamical problem where the system is prepared in a state $\phi=0$ at
the time $t=0$ (this could have been by cooling the system to a very low temperature for instance)
and then letting it relax at some non-zero temperature $T$. We will consider the very general relaxational dynamics, where the evolution of the field is given by
\begin{equation}
{\partial \phi({\bf x})\over \partial t} =  -R\Delta \phi({\bf x},t) +\eta({\bf x,t})  \label{dyn}
\end{equation}
where the noise is Gaussian and  uncorrelated in time
mean with correlation function
$\langle \eta({\bf x},t )\eta({\bf x}',t')\rangle = 2T\delta(t-t')R({\bf x} -{\bf x}')$. 
Here $R$ is a symmetric translationally invariant operator and the choice of the noise correlator ensures
that the dynamics obeys detailed balance. For thermal noise uncorrelated in time this is a very general
representation of the dynamics of soft condensed matter systems, such as those mentioned above,  when intertial and relativistic effects (due to the presence of large viscosities) can be neglected.
The formal solution to this equation, for our initial condition,   is
\begin{equation}
\phi({\bf x},t) = \int_0^t ds \exp\left(-(t-s)R\Delta\right)\eta({\bf x},s)
\end{equation}
This means that the equal time correlation function of the field, $C({\bf x}, {\bf x}',t)=\langle \phi({\bf x},t)\phi({\bf x}',t)\rangle$, can be shown to be given by
\begin{equation}
C(t) = TR\left[ 1- \exp(-2t\Delta R)\right] (\Delta R)^{-1} \label{corrfn}
\end{equation}

Now if we Laplace transform this equation (defining $ {\cal L} f(s) = \int_0^\infty dt \exp(-st) f(t)$)we find that
\begin{equation}
{\cal L} C({\bf x},{\bf x}',s) = {T\over s} \Delta^{-1}_s({\bf x}, {\bf x},l')
\end{equation}
where 
\begin{equation}
\Delta_s = \Delta  +{sR^{-1}\over 2}\label{eqds}
\end{equation}
  \begin{figure}
\begin{center}
\resizebox{8 cm}{!}{\includegraphics{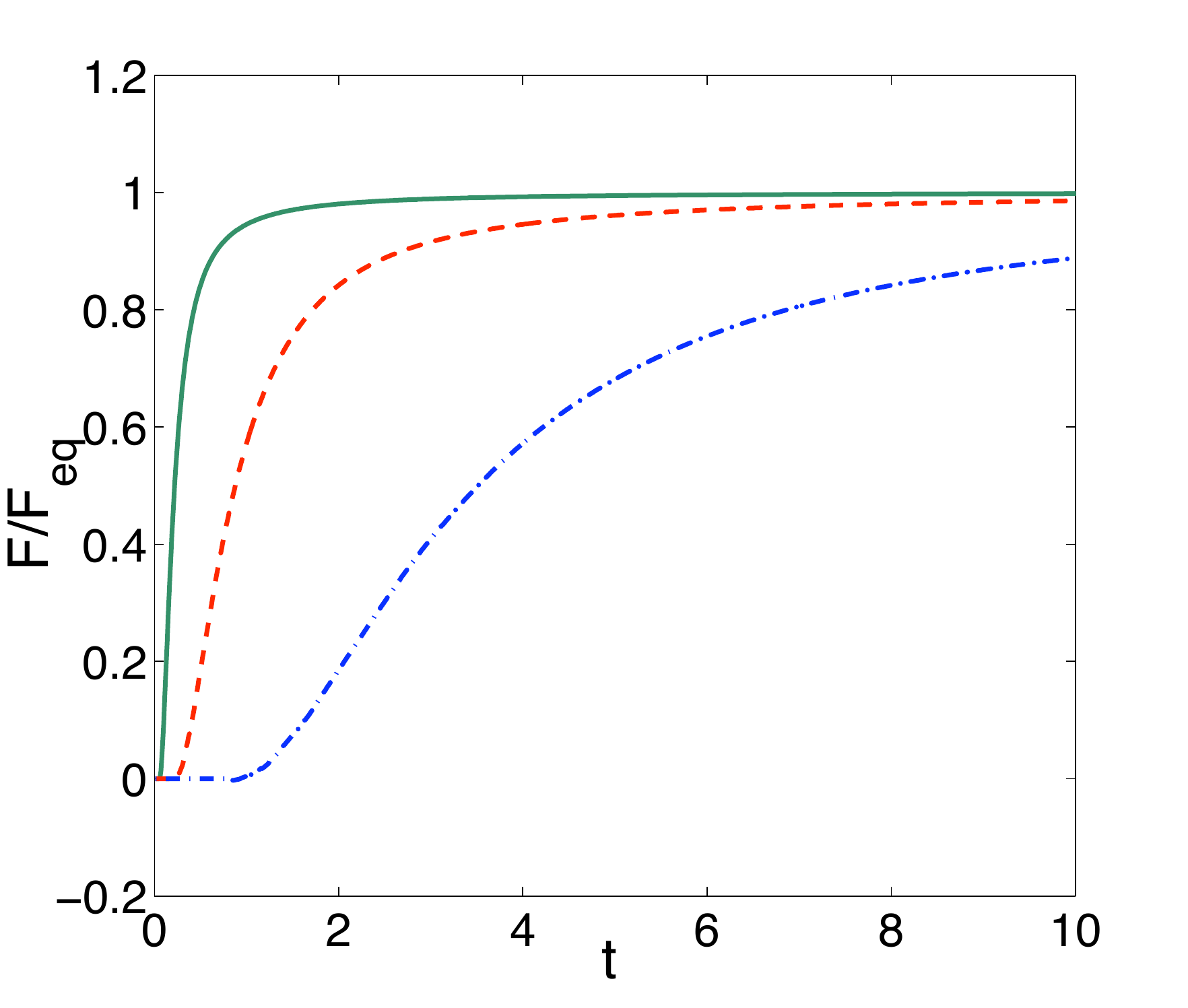}}
\end{center}
\vspace*{-.6cm}
\caption{ Approach to equilibrium for various values of the plate separation $l$ (top to bottom $l=1,2,4$) for $d=3$. Nonequilibrium  force $F$, obtained by the direct numerical inverse Laplace transform of Eq.~(\ref{laplaceF}), plotted in units of the equilibrium pseudo-Casimir force ($F_{eq}$) as a function of time.}
\vspace*{-.6cm}
\label{fig1}
\end{figure}
Now using this and Eq.(\ref{eqforce}) we find that the Laplace transform of the average value of the generalised force is given by
\begin{equation} 
\langle {\cal L} F_l(s)\rangle= -{T\over 2s} 
\int d{\bf x} d{\bf x}' {\partial \over \partial l} \Delta({\bf x}, {\bf x}', l)\Delta^{-1}_s({\bf x}, {\bf x}'),\label{lt1}
\end{equation}
which using Eq, (\ref{eqds}) and Eq. ({\ref{eqfe}) can be written as 
\begin{eqnarray}
\langle {\cal L} F_l(s)\rangle &=& -{T\over 2s} 
\int d{\bf x} d{\bf x}' {\partial \over \partial l} \Delta_s({\bf x}, {\bf x}', l)\Delta^{-1}_s({\bf x}, {\bf x}') \nonumber \\
&=&  {T\over s}{\partial\over \partial l}\ln(Z(\Delta_s)) 
\label{lap-partition}
\end{eqnarray}
This result is quite remarkable. It means that the Laplace transform of the time dependent pseudo-Casimir force considered here is given by a static pseudo-Casimir force with the operator $\Delta_s$. It is clear that the result is also valid for the force  on any surface in the system
which interacts with the field. Providing the static partition function is known for the  screened problem, the corresponding time dependent force can be extracted by inverting the Laplace transform.  

We now turn to the case where the imposed boundary conditions are Dirichlet and the two plates are immersed in the fluctuating medium (hence there is fluctuating medium on both sides of each plate).
For the relaxational dynamics of the field (Eq.\ref{dyn}) we will consider the simplest case with $R=
\delta({\bf x}-{\bf x}')$, which means that the Gaussian noise is both uncorrelated in space and time. This corresponds to dynamics that does not conserve the order parameter and is also known as model A dynamics. 
We take the total length of the system to be $L$ which is fixed, and place the plates of area $A$ 
a distance $l$ apart.   
Standard results on the screened pseudo-Casimir interaction \cite{kar1999} give
\begin{equation}
\langle {\cal L} F_l(s)\rangle = -{2AT\over s (4\pi)^{d-1\over 2} \Gamma({d-1\over2})} I(l,s), 
\label{laplaceF}
\end{equation}
where $d$ is the dimension of the space, $A$ is the area of the plates and,
 \begin{equation}
 I(l,s) =  \int k^{d-2} dk  {\sqrt{k^2 + {s\over 2}}\exp(-2l\sqrt{k^2 + {s\over 2}})\over 1-\exp(-2l\sqrt{k^2 + {s\over 2}})},
\end{equation}
 The equilibrium behavior is easily extracted by examining the pole at $s=0$ which yields, as anticipated, the standard equilibrium pseudo-Casimir force \cite{kre1992}
\begin{equation}
\langle F_l\rangle_{eq} =-{AT\Gamma(d)\zeta(d)\over (16\pi)^{{d-1}\over 2}
  \Gamma({d-1\over 2}) l^d},
\end{equation}
where $\Gamma$ is Euler's gamma function and $\zeta$ is the Riemann zeta function \cite{gra2000}. The full time dependence of the force, starting at zero at $t=0$ and relaxing to the equilibrium value above, can be extracted by direct Laplace inversion of Eq. (\ref{laplaceF}). Fig.(\ref{fig1}) shows the approach to equilibrium for three different plate separations. Clearly, the relaxation times increase with plate separation and this is due to the fact that the underlying dynamics is diffusive and hence $l^2$ sets a time scale. Useful {\it analytic} expressions for the early and late time behavior of the non-equilibrium force can also be obtained from Eq.(\ref{laplaceF}). Using specific properties of Laplace transforms and the fact that 
$F_l(t=0)=0$ allows us to arrive at an expression for the time derivative of the  force

\begin{equation}
\langle {dF_l\over dt} \rangle =-{AT\over (8\pi)^{d\over 2} t^{d+2\over 2}}\sum_{n=1}^\infty 
\left[{l^2 n^2\over t} - 1\right]\exp(-{l^2 n^2\over 2 t}), 
\end{equation}
which in turn gives us the asymptotic behavior of the pseudo-Casimir force for different time regimes defined by  $l^2/t $ 
\begin{eqnarray}
\langle F_l(t)\rangle  &\sim&  -{2AT\over (8\pi t )^{d\over 2} }\exp(-{l^2\over 2t}) \ \ {\rm for} \ \ {l^2\over t} \gg 1  \nonumber \\
&\sim& \langle F_l\rangle_{eq} + {AT\over d(8\pi)^{d\over 2} t^{d\over 2}} \ \ {\rm for} \ \ {l^2\over t} \ll 1 
\label{etlt}
\end{eqnarray}
 It is interesting to note that the late time correction is independent of $l$. This is because the medium between the two plates has a relaxation time $\tau(l) \sim l^2/2\pi^2$ whereas the slowest relaxation times
in the system are associated with the medium outside the two plates and hence at late times the correction is dominated by the relaxation of the external system in the thermodynamic limit $L\to \infty$. This diffusive relaxation is responsible for the  power law approach to equilibrium. It is natural to ask how these model 
A results translate into the stress tensor formalism. Because the rhs of Eq. (\ref{lap-partition})
is static we can write the result in terms of the stress tensor for the corresponding field theory, and inverting the Laplace transform we find  \cite{inprep} an effective dynamical stress tensor:
\begin{equation}
T^{dyn}_{ij} = {1\over 2}\delta_{ij}\left[[\nabla\phi]^2 +{1\over 2}{\partial \over \partial t }\phi^2\right] -\nabla_i\phi\nabla_j\phi.
\end{equation}  
Thus for the problem considered here we see that the first part of the stress tensor picks up a time derivative term whose expectation will vanish in equilibrium to give the usual equilibrium stress tensor result. 
It is also clear that simply using the standard stress tensor on a surface in a system out of equilibrium
will only give the right result when Dirichlet boundary conditions are imposed (as $\phi=0$ on the surface).
 
The method can also be used to examine the dynamics resulting from a sudden change in temperature, from say $T_0$ where the system is in equilibrium to a temperature $T$. In this case, the initial configuration of the field $\phi({\bf x},0)$ has the correlation function
\begin{equation}
\langle \phi({\bf x},0)\phi({\bf x}',0)\rangle = T_0\Delta^{-1}({\bf x},{\bf x'},l).
\end{equation}
Solving the equation of motion Eq.(\ref{dyn}) with this initial condition yields the time
dependent correlation function
\begin{equation}
 C({\bf x},{\bf x}',t)= T_0 \exp(-2t \Delta)\Delta^{-1} + T (1-\exp(-2t \Delta))\Delta^{-1}
\end{equation}
The same reasoning as for Eq.(\ref{corrfn})-(\ref{lap-partition}) yields
\begin{equation}
\langle {\cal L} F_l(s)\rangle =  {T_0 \over s}{\partial\over \partial l}\ln(Z(\Delta)) 
+{T-T_0\over s}{\partial\over \partial l}\ln(Z(\Delta_s)) 
\end{equation}
where we have used the fact that ${\partial\over \partial l}\ln(Z(\Delta_s))$ is independent of the 
temperature. For Dirichlet boundary conditions this gives, in analogy with Eq.(\ref{etlt}), the  limiting behavior
\begin{eqnarray}
\langle F_l(t)\rangle  &\sim&  \langle F_l\rangle_{eq\ T_0} -{2A(T-T_0)\over (8\pi t )^{d\over 2} }\exp(-{l^2\over 2t}) \ \ {\rm for} \ \ {l^2\over t} \gg 1  \nonumber \\
&\sim& \langle F_l\rangle_{eq\ T} +{A(T-T_0)\over d(8\pi)^{d\over 2} t^{d\over 2}} \ \ {\rm for} \ \ {l^2\over t} \ll 1 
\end{eqnarray}

One can also consider the behavior of the pseudo-Casimir force for relaxational dynamics of the form
of Eq.(\ref{dyn}) but where the forcing noise is {\it colored} in time such that
$\langle \eta( {\bf x},t)\eta({\bf x}',t')\rangle = T\delta({\bf x}-{\bf x}') \omega\exp(-\omega |t-t'|)$ (so $R({\bf x}-{\bf x}')=\delta({\bf x}-{\bf x}')$ has the model A from). Here 
$T$ represents an energy scale, $\omega$ a frequency and the resulting steady state is not an equilibrium one. The average 
value of the force in the steady state regime can be computed using the same formalism above and we
find that the correlation function of the field is given by
\begin{equation}
C({\bf x},{\bf x'}, \omega ) = T\left[ \Delta({\bf x},{\bf x'}, l)^{-1} - \Delta_{2\omega}({\bf x},{\bf x'}, l)^{-1}\right]
\end{equation}
which yields
\begin{equation}
\langle F_l(\omega)\rangle = {T}{\partial\over \partial l}\left[\ln(Z(\Delta)) -\ln(Z(\Delta_{2\omega})\right] \label{coln}
\end{equation}
Hence again we find that one can compute a force in a non equilibrium system from  knowledge of static screened systems \cite{comment}. Note that in the limit $\omega \to \infty$ we recover the white noise equilibrium  result of Eq. ({\ref{eqfe}). Fig. (\ref{fig3}) shows the frequency dependence of the non-equilibrium force obtained from Eq.(\ref{coln}) for a two plate system with Dirichlet boundary conditions as we had before. Again $l^2$ sets a timescale and we see that for $\omega \gg l^{-2}$ the force, $F$ tends to the equilibrium white noise value, $F_{eq}$, as expected, while for $\omega \ll l^{-2}$, $F \ll F_{eq}$ and as $\omega \to 0$, $F $ vanishes. The inset to Fig. (\ref{fig3}) shows how the force depends on plate separation for fixed $\omega$. Again, equilibrium behavior is recovered for large $l$ ($\omega \gg l^{-2}$), while for small plate separations the force changes qualitatively scaling as $l^{-1}$. We note that the result Eq.(\ref{coln}) agrees with a computation for the same system where the steady state  force was computed using the stress tensor \cite{ba2003}. We finally note that our approach can also be readily extended to study the non-equilibrium pseudo-Casimir force between two small {\em defect} regions within the pairwise approximation  \cite{inprep}.
\begin{figure}
\begin{center}
\resizebox{8 cm}{!}{\includegraphics{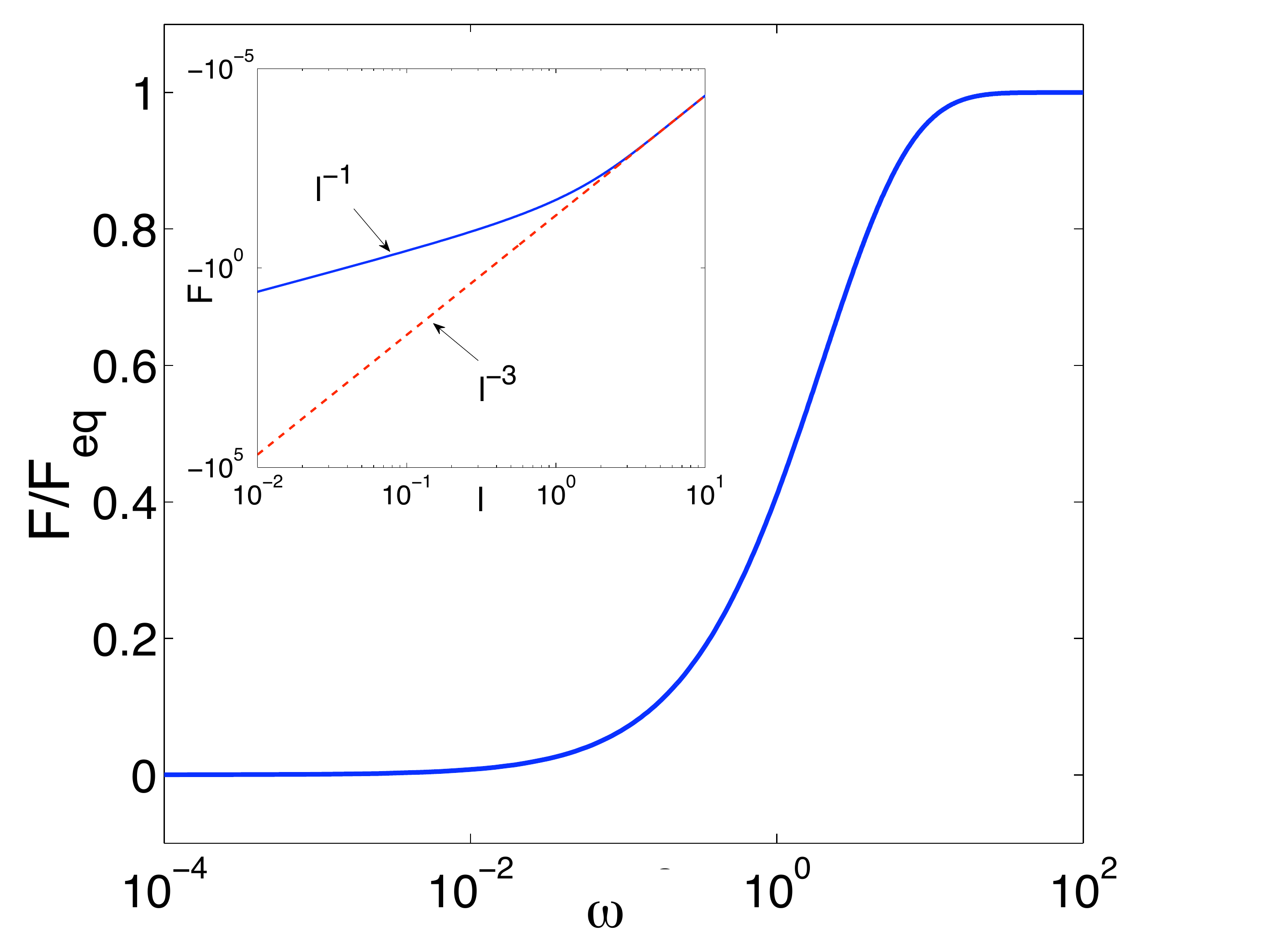}}
\end{center}
\vspace*{-.6cm}
\caption{ Nonequilibrium pseudo-Casimir force, $F$, in units of the equilibrium force ($F_{eq}$) as a function of $\omega$ for $l=1$ and $d=3$. Inset: $F$,  as a function of plate separation $l$ for $\omega=1$ and $d=3$ (solid line). The equilibrium force (dashed line) is shown for comparison.}
\vspace*{-.6cm}
\label{fig3}
\end{figure}

 Previous studies on the dynamical pseudo-Casimir effect concentrated on steady state non equilibrium dynamics or dynamics close to equilibrium and considered Dirichlet boundary conditions assuming that the equilibrium stress tensor could be applied to compute the force. Our formalism marks a major advance that overcomes these restrictions by allowing the time-dependent force to be evaluated 
unambiguously via an expression for the energy of the field. We emphasize the generality of our approach
as it is valid for (i) any dissipative dynamics of the form Eq. (\ref{dyn}) , (ii)  any free field theory ({\em e.g.} operators $\Delta$ which contain
terms such as $\nabla^4$ as is the case for the height fluctuations of lipid membranes) and (iii) for any 
generalized force conjugate to any parameter in the system. It is worth noticing that the 
static theory that must be solved for conserved dynamics (model B), where $R=-\nabla^2$, becomes
non-local and the corresponding static calculation presents us with the interesting problem of understanding
pseudo-Casimir forces for systems with non local interactions.     
 This research was supported in part by the National Science Foundation under Grant No.PHY05-51164 (while at the KITP UCSB  program
{\em The theory and practice of fluctuation induced interactions} 2008). DSD acknowledges support from   the Institut Universitaire de France.

\end{document}